\documentclass[conference]{IEEEtran}

\usepackage[linesnumbered,lined, algoruled]{algorithm2e}
\usepackage{amssymb,bm}
\usepackage{amsmath}
\usepackage[acronym,shortcuts]{glossaries}
\usepackage{algorithmic,float}
\usepackage{balance}
\usepackage{color}
\usepackage{dsfont}
\usepackage{soul}
\usepackage{graphicx}
\usepackage{comment}
\usepackage{mathtools}
\usepackage{placeins}

\graphicspath{{./figures/}}

\makeatletter
\newsavebox\myboxA
\newsavebox\myboxB
\newlength\mylenA
\newcommand*\xoverline[2][0.75]{%
	\sbox{\myboxA}{$\m@th#2$}%
	\setbox\myboxB\null
	\ht\myboxB=\ht\myboxA%
	\dp\myboxB=\dp\myboxA%
	\wd\myboxB=#1\wd\myboxA
	\sbox\myboxB{$\m@th\overline{\copy\myboxB}$}
	\setlength\mylenA{\the\wd\myboxA}
	\addtolength\mylenA{-\the\wd\myboxB}%
	\ifdim\wd\myboxB<\wd\myboxA%
	\rlap{\hskip 0.5\mylenA\usebox\myboxB}{\usebox\myboxA}%
	\else
	\hskip -0.5\mylenA\rlap{\usebox\myboxA}{\hskip 0.5\mylenA\usebox\myboxB}%
	\fi}
\makeatother

\makeglossaries
\newacronym{3gpp}{3GPP}{3rd Generation Partnership Project}
\newacronym{5g}{5G}{fifth generation}
\newacronym{6g}{6G}{sixth generation}
\newacronym{ap}{AP}{access point}
\newacronym{awgn}{AWGN}{additive white Gaussian noise}
\newacronym{bler}{BLER}{block error rate}
\newacronym{bs}{BS}{base station}
\newacronym{cdf}{CDF}{cumulative distribution function}
\newacronym{csi}{CSI}{channel state information}
\newacronym{ecf}{ECF}{empirical characteristic function}
\newacronym{eesm}{EESM}{exponential effective SINR metric}
\newacronym{fa}{FA}{false alarm}
\newacronym{glrt}{GLRT}{generalized likelihood ratio test}
\newacronym{ini}{InI}{indoor industrial}
\newacronym{inm}{InM}{indoor mixed}
\newacronym{ino}{InO}{indoor office}
\newacronym{iad}{IAD}{inter-AP distance}
\newacronym{kpi}{KPI}{key performance indicator}
\newacronym{jr}{JR}{joint reception}
\newacronym{los}{LOS}{line of sight}
\newacronym{lte}{LTE}{long term evolution}
\newacronym{mrc}{MRC}{maximum ratio combining}
\newacronym{mbb}{MBB}{mobile broadband}
\newacronym{md}{MD}{missed detection}
\newacronym{mimo}{MIMO}{multiple-input multiple-output}
\newacronym{mle}{MLE}{maximum likelihood estimate}
\newacronym{mse}{MSE}{mean squared error}
\newacronym{mmse}{MMSE}{minimum mean squared error}
\newacronym{np}{NP}{Neyman-Pearson}
\newacronym{nr}{NR}{new radio}
\newacronym{ofdm}{OFDM}{orthogonal frequency division multiplexing}
\newacronym{pdcp}{PDCP}{packet data convergence protocol}
\newacronym{pdf}{PDF}{probability density function}
\newacronym{pdr}{PDR}{packet delivery ratio}
\newacronym{pmf}{PMF}{probability mass function}
\newacronym{prb}{PRB}{physical resource block}
\newacronym{re}{RE}{resource element}
\newacronym{rf}{RF}{radio frequency}
\newacronym{roc}{ROC}{receiver operating characteristic}
\newacronym{rlrt}{RLRT}{Roy's largest root test}
\newacronym{rss}{RSS}{received signal strength}
\newacronym{rv}{r.v.}{random variable}
\newacronym{sar}{SAR}{single access point reception}
\newacronym{se}{SE}{spectral efficiency}
\newacronym{sinr}{SINR}{signal to interference plus noise ratio}
\newacronym{siso}{SISO}{single-input single-output}
\newacronym{snr}{SNR}{signal to noise ratio}
\newacronym{tdd}{TDD}{time division duplex}
\newacronym{ue}{UE}{user equipment}
\newacronym{ul}{UL}{uplink}
\newacronym{urllc}{URLLC}{ultra-reliable low-latency communications}

\begin{document}

\title{Jamming Resilient Indoor Factory Deployments: Design and Performance Evaluation}

\author{Leonardo Chiarello$^{*\mathsection}$, Paolo Baracca$^\mathsection$, Karthik Upadhya$^\dagger$,\\
	 Saeed R. Khosravirad$^\ddagger$, Silvio Mandelli$^\mathsection$, and Thorsten Wild$^\mathsection$\\
$^*$Department of Information Engineering, University of Padova, Italy\\
$^\mathsection$Nokia Bell Labs, Stuttgart, Germany\\
$^\dagger$Nokia Bell Labs, Espoo, Finland\\
$^\ddagger$Nokia Bell Labs, Murray Hill, USA\\}

\maketitle

\sloppy
\begin{abstract}
	In the framework of 5G-and-beyond Industry 4.0, jamming attacks for denial of service are a rising threat which can severely compromise the system performance. Therefore, in this paper we deal with the problem of jamming detection and mitigation in indoor factory deployments. We design two jamming detectors based on pseudo-random blanking of subcarriers with orthogonal frequency division multiplexing and consider jamming mitigation with frequency hopping and random scheduling of the user equipments. We then evaluate the performance of the system in terms of achievable \ac{bler} with ultra-reliable low-latency communications traffic and jamming missed detection probability. Simulations are performed considering a 3rd Generation Partnership Project spatial channel model for the factory floor with a jammer stationed outside the plant trying to disrupt the communication inside the factory. Numerical results show that jamming resiliency increases when using a distributed access point deployment and exploiting channel correlation among antennas for jamming detection, while frequency hopping is helpful in jamming mitigation only for strict \ac{bler} requirements.
\end{abstract}


\glsresetall

\section{Introduction}\label{sec:intro}

Security has been one of the main drivers in the design of the \ac{5g} of mobile communication systems by the \ac{3gpp}. In fact, \ac{5g} provides several security measures at higher layers to guarantee authentication, privacy and data integrity \cite{dahlman_2018}. Moreover, radio jamming by a malicious device has also been recognized as an important type of security attack that can threaten the performance of a \ac{5g} deployment, in particular in Industry 4.0 scenarios.
Despite the very affordable cost with a starting price of a few hundred dollars \cite{jammerstore_2021}, some of these devices can be quite advanced and smart, e.g., the so-called reactive jammers \cite{wilhelm_2011}, as capable to sense the channel and remain quiet until an ongoing legitimate transmission is detected. In fact, \ac{urllc} are inherently more susceptible to the interference impact of such a denial of service attack due to their stringent quality of service requirements. For instance, a jammer stationed outside a factory that disrupts the communication among the devices inside the plant can cause large economic losses to the factory owner if production needs to be stopped. Furthermore, handling jamming attacks has already been recognized as a very relevant theme  also for \ac{6g} technologies \cite{berardinelli_2021}, with physical layer security expected to play an important role in future mobile networks \cite{chorti_2021}.


A jamming resilient communication system must provide both a) detection, to discriminate between the presence of a jammer and legitimate interference, and b) mitigation capabilities, to limit the caused damage by applying ad-hoc techniques.
Non-reactive jammers can be detected by monitoring basic statistics like the received signal strength or the carrier sensing time, whereas the detection of smart jammers require advanced techniques combining several statistics \cite{xu_2006}. In \cite{do_2018} authors propose a detection technique based on pseudo-random hopping of the scheduled \acp{ue} among the pilot sequences and the application of a jamming-resilient combiner exploiting massive \ac{mimo} base stations. In our previous work \cite{chiarello_2021}, we proposed a novel method to detect smart jamming attacks based on pseudo-random blanking of subcarriers with \ac{ofdm}. Regarding the mitigation problem, several schemes have already been studied, for instance applying beamforming, direct sequence spread spectrum, and power control \cite{grover2014jamming}. In fact, once a jammer is detected and characterized, an off-the-shelf interference management scheme can be applied tailoring it to the specific scenario, e.g., with beamforming creating  nulls toward a jammer whose channel can be estimated in the detection phase.

In this paper we extend the jamming detection proposal in \cite{chiarello_2021} by providing realistic performance evaluations that consider indoor factory deployments with \ac{3gpp} spatial channel model. Moreover, we propose a new detector that exploits antenna correlation at the receiver. Finally, we consider jamming mitigation techniques with frequency hopping and random scheduling of the \acp{ue}. The benefits of the proposed schemes are evaluated in terms of jamming detection probability and \ac{bler} performance with \ac{urllc}.

\emph{Notation}. We use $(\cdot)^{\textnormal{H}}$ to denote conjugate transpose. $\lVert \mathbf{x} \rVert$ indicates the norm of vector $\mathbf{x}$. $\lvert \cdot \rvert$ denotes the absolute value. $[ \mathbf{x} ]_n$ is the $n$-th entry of vector $\mathbf{x}$. $F_X^{-1}(x)$ denotes the inverse of the \ac{cdf} of the \ac{rv} $X$ evaluated at $x$.
\section{System Model}\label{sec:system_model}

\begin{figure}
	\centering
	\includegraphics[width=.7\columnwidth]{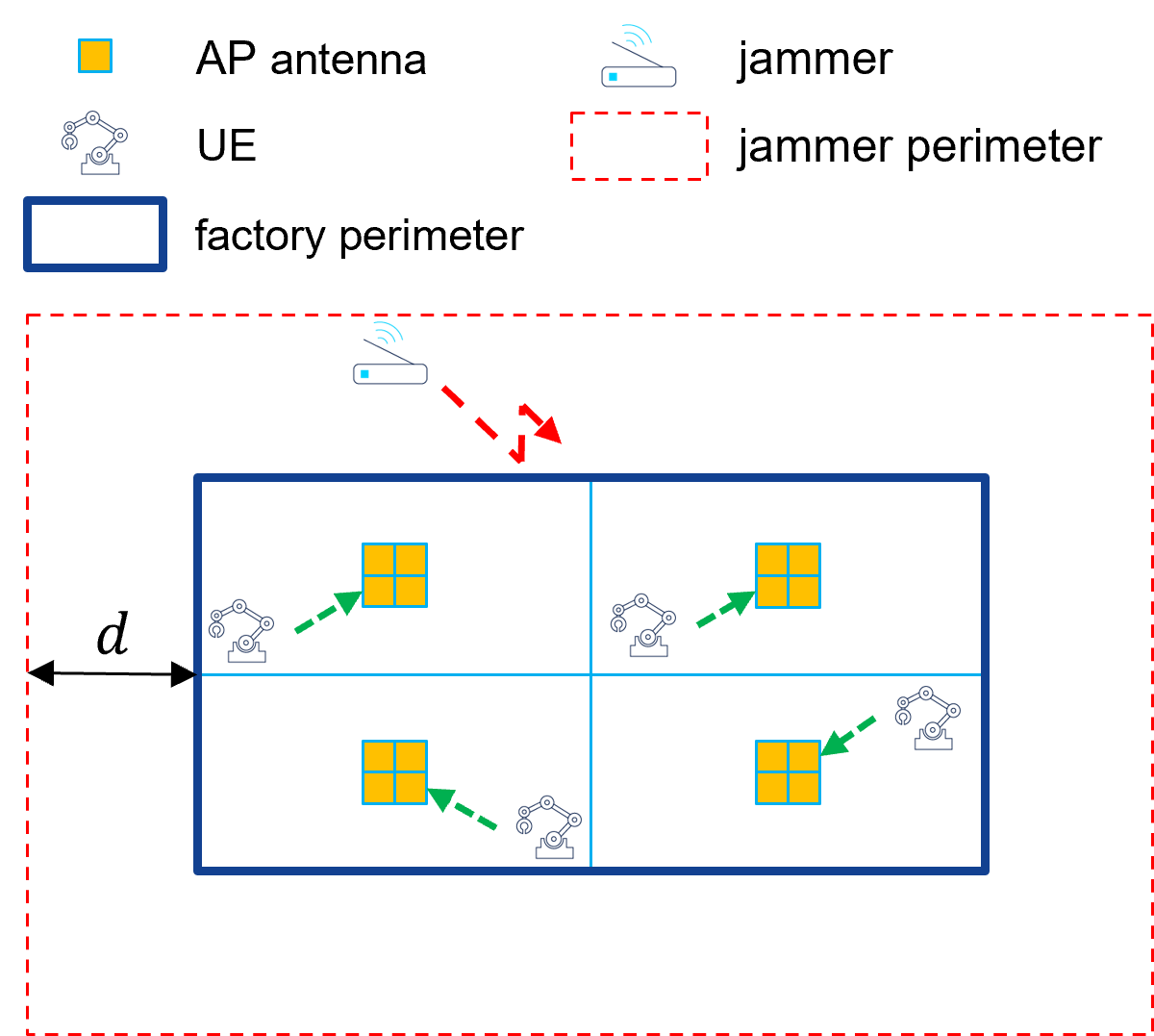}
	\caption{Representation of the considered uplink scenario for the partially distributed deployment ($N_{\textnormal{AP}}=4$) and a total of $N_{\textnormal{ant}} = 16$ antennas.}
	\label{fig:scenario}
\end{figure}

We consider an industrial scenario as in Fig. \ref{fig:scenario} with a factory hall of dimensions $100 \times 50 \times 6 \textnormal{ m}$, and with $N_{\textnormal{AP}}$ \acp{ap} mounted on the factory ceiling. For a fair comparison among different deployments, we consider in the whole factory a total of $N_{\textnormal{ant}}$ omni-directional antennas so that each \ac{ap} is equipped with a square antenna array with $N_{\textnormal{ant}}^{(\textnormal{AP})} = N_{\textnormal{ant}} / N_{\textnormal{AP}}$ antennas. The following \ac{ap} deployments are compared \cite{alonzo2021cell}:
\begin{itemize}
	\item \emph{Centralized deployment}: $N_{\textnormal{AP}}=1$ \ac{ap} placed at the center of the factory hall;
	\item \emph{Partially distributed deployment}: $N_{\textnormal{AP}}=4$ \acp{ap} located such that the \ac{iad} along the longest side is $50 \, \textnormal{m}$ and the \ac{iad} along the shortest side is $25 \, \textnormal{m}$. An example of this deployment is reported in Fig. \ref{fig:scenario}.
	\item \emph{Fully distributed deployment}: $N_{\textnormal{AP}}=16$ \acp{ap} located such that the \ac{iad} along the longest side is $25 \, \textnormal{m}$ and the \ac{iad} along the shortest side is $12.5 \, \textnormal{m}$.
\end{itemize}
We have $N_{\textnormal{UE}}$ \acp{ue} active and each \ac{ue} is randomly dropped within the factory at an height of $1.5 \, \textnormal{m}$, is equipped with a single omni-directional antenna, and transmits with power $P_{\textnormal{UE}} = 10 \, \textnormal{dBm}$.

We assume a system operating at a central carrier frequency of $f_\textnormal{C} = 3.75 \, \textnormal{GHz}$. Regarding the channel model, we consider the proposal in \cite{R1-1813177}, where the \ac{3gpp} \ac{ino} model is used as starting point and  path-loss, shadowing, and \ac{los} probability values are chosen on the basis of extensive measurements done in two different operational factories.
This novel \ac{ini} model encompasses different scenarios and here we consider the dense factory clutter model with clutter-embedded APs (more details in \cite[Tab. 3]{R1-1813177}).

\subsection{Numerology and resource allocation}\label{subsec:numerology}

We adopt an \ac{ofdm} modulation compliant to the \ac{5g} numerology with $60 \, \textnormal{kHz}$ subcarrier spacing. The subcarriers are grouped into \acp{prb}, each consisting of $N_{\textnormal{sc}} = 12$ consecutive subcarriers over a transmission interval of $N_{\textnormal{symb}} = 14$ \ac{ofdm} symbols \cite{dahlman_2018}. Therefore, each \ac{prb} consists of $N_{\textnormal{RE}}^{(\textnormal{PRB})} = N_{\textnormal{sc}} \cdot N_{\textnormal{symb}} = 168$ \acp{re} and has a bandwidth of $B_{\textnormal{PRB}} = 720 \, \textnormal{kHz}$. We consider two scenarios for our system: a total bandwidth of $B =  20 \, \textnormal{MHz}$ (with a total number of \acp{prb} $N_{\textnormal{PRB}} = 25$) with $N_{\textnormal{UE}}=4$ \acp{ue}, and a total bandwidth of $B =  100 \, \textnormal{MHz}$ ($N_{\textnormal{PRB}} = 125$) with $N_{\textnormal{UE}}=20$ \acp{ue}: in both cases we set the guard band to be 10\% of $B$.
We assume \ac{urllc} traffic, such that each \ac{ue} transmits a small packet of size $C = 20 \, \textnormal{bytes}$ in each slot, with no retransmission opportunities because of the tight latency constraint. We consider a resource allocation where interference among the active \acp{ue} is managed by allocating different \acp{ue} on different \acp{prb}, i.e., the only interference source in the system is the jammer. The \acp{prb} available for data transmissions are then evenly shared among the \acp{ue}, that apply equal power allocation on them. More details about the allocation of \acp{ue} to \acp{prb} is part of the jamming mitigation strategy and will be described in Section \ref{sec:defense_strategy}.

\subsection{Jammer model}

We consider an attacker stationed outside the factory at height of $1.5 \, \textnormal{m}$ and dropped randomly within a rectangular perimeter with sides $d = 10 \, \textnormal{m}$ far from the factory walls (see Fig. \ref{fig:scenario}). The jammer is equipped with a single omni-directional antenna element that transmits with power $P_{\textnormal{J}}$, ranging from $20 \, \textnormal{dBm}$ to $60 \, \textnormal{dBm}$ \cite{jammerstore_2021}. 
Moreover, we assume the jammer to allocate equal power on the attacked \acp{prb} and consider both a) a wide-band jammer that attacks the whole bandwidth and b) a narrow-band jammer attacking a few \acp{prb} but with stronger power spectral density.
Finally, we assume for the jammer the same \ac{ini} channel model as for the \acp{ue} inside the factory, but adding a factory wall penetration loss modelled as a Gaussian \ac{rv} $\textnormal{PL}_\textnormal{wall} \sim \mathcal{N}(\mu_P,\sigma_P^2)$, with mean  $\mu_P = 27.5 \, \textnormal{dB}$ and standard deviation $\sigma_P = 6.5 \, \textnormal{dB}$ \cite[Tab. 7.4.3-2]{38901}.

\subsection{Imperfect \ac{csi}}

We assume a \ac{tdd} setup with pilot sequence length $T=16$ \cite{alonzo2021cell}. Note that here we have no pilot contamination as different \acp{ue} are scheduled on different \acp{prb}, but jamming affects channel estimation. Let $\mathbf{h}_{i,j}$ be the $(1 \times N_{\textnormal{ant}}^{(\textnormal{AP})})$-dimensional channel vector from the $i$-th \ac{ue} to the $j$-th \ac{ap} on a certain \ac{prb}, with $i = 1, \dots, N_{\textnormal{UE}}$, and $j = 1, \dots, N_{\textnormal{AP}}$. The \ac{mmse} estimate $[\widehat{\mathbf{h}}_{i,j}]_n$ of $[\mathbf{h}_{i,j}]_n$ can be defined as \cite[Eq. (6)]{alonzo2021cell}:
\begin{equation}
	\label{eq:chan_est}
	[\widehat{\mathbf{h}}_{i,j}]_n = \frac{\gamma_{i,j} T}{1 + \gamma_{i,j} T} \left( [\mathbf{h}_{i,j}]_n + z_i \right) \, ,
\end{equation}
where  $\gamma_{i,j} = P_{\textnormal{UE},i}^{(\textnormal{PRB})} \sigma_{h_{i,j}}^2 / \sigma_w^2$ is the \ac{snr} of \ac{ue} $i$ at \ac{ap} $j$ and $z_i \sim \mathcal{CN} ( 0, (\sigma_w^2 + P_{\textnormal{J}}^{(\textnormal{PRB})} \sigma_{h_{\textnormal{J},j}}^2) / (P_{\textnormal{UE},i}^{(\textnormal{PRB})} T) )$ is a complex Gaussian \ac{rv} representing noise and interference on channel estimation. In particular, $P_{\textnormal{UE},i}^{(\textnormal{PRB})}$ is the power of \ac{ue} $i$ allocated to a single \ac{prb}, $\sigma_{h_{i,j}}^2$ denotes the large-scale fading attenuation between \ac{ue} $i$ and \ac{ap} $j$, and $\sigma_w^2$ is the noise statistical power on a single \ac{prb}, computed considering a noise figure of $7 \, \textnormal{dB}$ at the receiver. Moreover, $P_{\textnormal{J}}^{(\textnormal{PRB})}$ is the jammer power allocated to a single \ac{prb} and $\sigma_{h_{\textnormal{J},j}}^2$ is the large-scale fading attenuation between the jammer and \ac{ap} $j$. 

\subsection{Beamforming at the receiver}

At the receiver, we assume \ac{jr}, such that the signals received by the \acp{ap} are combined in a central unit. Since there is no interference among the active \acp{ue} in our framework, because they are scheduled on different subbands, we adopt \ac{mrc}, that maximizes the \ac{ue} \ac{snr} and is easy to implement in a distributed \ac{mimo} setup.
We denote with $\widehat{\mathbf{h}}_i = [ \widehat{\mathbf{h}}_{i,1}, \widehat{\mathbf{h}}_{i,2}, \dots, \widehat{\mathbf{h}}_{i,N_{\textnormal{AP}}} ]$ the $(1 \times N_{\textnormal{ant}})$-dimensional vector collecting the estimated channels between the $i$-th \ac{ue} and all the \acp{ap}. The \ac{mrc} beamforming is then defined as:
\begin{equation}
	\mathbf{g}_i = \widehat{\mathbf{h}}_i^{\textnormal{H}} / \lVert \widehat{\mathbf{h}}_i \rVert \, .
\end{equation}

\subsection{System \acp{kpi}}

In order to quantify the impact of the jammer to the system, we introduce two \acp{kpi}: \ac{sinr} on data transmission and \ac{bler}. We define the \ac{sinr} of \ac{ue} $i$ on a certain \ac{prb}, whose index is skipped for the sake of clarity, as
\begin{equation}
\label{eq:sinr}
	\textnormal{SINR}_i = \frac{\lvert \mathbf{h}_i \mathbf{g}_i \rvert^2 P_{\textnormal{UE},i}^{(\textnormal{PRB})}}{\sigma_w^2 + \lvert \mathbf{h}_\textnormal{J} \mathbf{g}_i \rvert^2 P_{\textnormal{J}}^{(\textnormal{PRB})}} \, ,
\end{equation}
where at the denominator we have the malicious interference from the jammer, with $\mathbf{h}_\textnormal{J}$ the $(1 \times N_{\textnormal{ant}})$-dimensional channel vector collecting the channels between the jammer and all the \ac{ap} antennas.

We assume that \ac{ue} $i$ sends its packet over $F_i$ \acp{prb} and define $C_{\textnormal{cod},i} = F_i \cdot N_{\textnormal{RE}}^{(\textnormal{PRB})}$ as the number of \acp{re} allocated to that packet. Then, for our analysis, we use the \ac{eesm} as link-to-system mapping criterion \cite[Eq. (3)]{brueninghaus_2005} to compute, as a function of the different \acp{sinr} (\ref{eq:sinr}) experienced by a certain \ac{ue} on different \acp{prb}, a single $\textnormal{SINR}_{\textnormal{pkt}}$, that represents the equivalent \ac{sinr} for the packet. We then use this $\textnormal{SINR}_{\textnormal{pkt}}$ to compute the \ac{bler} of \ac{ue} $i$ from the normal approximation of the finite blocklength capacity \cite[Eq. (5)]{durisi_2016}:
\begin{multline}
	\label{eq:bler}
	\textnormal{BLER}_{\textnormal{pkt},i} = Q \Bigg( \Bigg[ \log_2 \Bigg( 1 + \textnormal{SINR}_{\textnormal{pkt},i} \Bigg) - \rho_i \\
	+ \frac{\log_2 \widetilde{C}_{\textnormal{cod},i}}{2 \widetilde{C}_{\textnormal{cod},i}} \Bigg] \sqrt{\frac{\widetilde{C}_{\textnormal{cod},i}}{V}} \Bigg) \, ,
\end{multline}
where $V$ is the channel dispersion \cite[Eq. (8)]{durisi_2016}, $\rho_i = C / \widetilde{C}_{\textnormal{cod},i}$ is the spectral efficiency for the \ac{ue} $i$ packet, and $\widetilde{C}_{\textnormal{cod},i} = C_{\textnormal{cod},i} (1 - O)$ is the coded packet size in \acp{re} taking into account the system overhead $O=0.25$ for control and pilots.

\FloatBarrier
\section{Defense Strategy}\label{sec:defense_strategy}

In this work we consider the defense strategy framework for performing jamming detection that we initially proposed in \cite{chiarello_2021}, where some \acp{prb} in each slot are blanked in a pseudo-random manner, such that the attacker cannot predict in advance which resources will be used for transmission and which will be blanked. In detail, in each slot all the \acp{ue} blank a set $\mathcal{M}_P \subset \{ 1, \dots, N_{\textnormal{PRB}} \}$ (with cardinality $M_P = |\mathcal{M}_P|$) of \acp{prb}, where the set elements are chosen in a pseudo-random manner; the remaining \acp{prb} are used for data transmission. At the same time, the attacker transmits on a set $\mathcal{L}_P \subseteq \{ 1, \dots, N_{\textnormal{PRB}} \}$ (with cardinality $L_P = |\mathcal{L}_P|$) of \acp{prb}, where the set elements are chosen according to the jammer strategy. In this work we assume that the jammer chooses the attacked \acp{prb} pseudo-randomly and it evenly splits its power among them. Moreover, for the sake of notation, when $L_P = N_{\textnormal{PRB}}$ we refer to the attacker as a wide-band jammer, otherwise we call it narrow-band jammer. 

\subsection{Jamming detection strategies}

The detection strategy takes advantage of the blanked \acp{prb} to detect the presence of jamming by means of statistical hypothesis testing \cite{kay_1993}. Moreover, we assume that jamming detection is performed by a central unit collecting the signals received from all the \acp{ap} distributed in the factory hall. The two hypotheses for the sequence of blanked \acp{prb} are as follows:
\begin{itemize}
	\item There is no jamming and we have just thermal noise (null hypothesis $\mathcal{H}_0$);
	\item There is jamming (alternative hypothesis $\mathcal{H}_1$).
\end{itemize}
The above hypotheses translate to the following hypothesis test:
\begin{equation}
\label{eq:hyp_test}
	\begin{cases}
		\mathcal{H}_0: \mathbf{r} = \mathbf{w}\\
		\mathcal{H}_1: \mathbf{r} = \mathbf{w} + \mathbf{j}
	\end{cases} \, ,
\end{equation}
where $\mathbf{r}$, $\mathbf{w}$, and $\mathbf{j}$ are $((N_{\textnormal{RE}} \cdot N_{\textnormal{ant}}) \times 1)$-dimensional vectors,  with $N_{\textnormal{RE}} = M_P \cdot N_{\textnormal{RE}}^{(\textnormal{PRB})}$, containing the samples of the blanked \acp{re} of all the antennas. In particular, $\mathbf{r}$ is the total received signal by the \acp{ap}, $\mathbf{w}$ is the noise vector with elements $[\mathbf{w}]_n \sim \mathcal{CN} (0, \sigma_w^2 / N_{\textnormal{RE}}^{(\textnormal{PRB})})$, and $\mathbf{j}$ is the jamming signal with unknown distribution. Then, the test decides for $\mathcal{H}_1$ if
\begin{equation}
	T(\mathbf{r}) > \delta \, ,
\end{equation}
where $T(\mathbf{r})$ is the test statistic and $\delta$ is the threshold, which depends on the test statistic and is function of a target \ac{fa} probability $P_{\textnormal{FA}}$, i.e., the probability of declaring jamming even if it is not present. Then, in Section \ref{sec:sim_results} we will evaluate the effectiveness of the proposed detection technique against a Gaussian jammer in terms of \ac{md} probability $P_{\textnormal{MD}}$, i.e., the probability of declaring no-jamming even if it is present. Note that with (\ref{eq:hyp_test}) we perform jamming detection in each slot: however, the proposed scheme can be applied, depending on the use case, also to multiple slots for improved performance. Regarding the test statistic, we now propose two options.

\subsubsection{\Ac{glrt}}

This test defines the test statistic simply as \cite{chiarello_2021}
\begin{equation}
\label{eq:glrt}
	T_{\textnormal{GLRT}} = \frac{\lVert \mathbf{r} \rVert^2}{N_{\textnormal{RE}} \cdot N_{\textnormal{ant}}} \, ,
\end{equation}
which is an energy detector. The threshold for this detector is derived as
\begin{equation}
	\delta_{\textnormal{GLRT}} = F^{-1}_{T_{\textnormal{GLRT}} (\mathbf{r}; \mathcal{H}_0)} (1 - P_{\textnormal{FA}}) \, ,
\end{equation}
where $T_{\textnormal{GLRT}} (\mathbf{r}; \mathcal{H}_0) \sim \textnormal{Gamma} \left( N_{\textnormal{RE}} \cdot N_{\textnormal{ant}}, \frac{\sigma_w^2 / N_{\textnormal{RE}}^{(\textnormal{PRB})}}{N_{\textnormal{RE}} \cdot N_{\textnormal{ant}}} \right)$ is the test statistic distribution under $\mathcal{H}_0$, with $\textnormal{Gamma}(k, \theta)$ being the gamma distribution with shape parameter $k$ and scale parameter $\theta$. The main advantage of this detector is the very low computational complexity, as just the received power on the blanked \acp{prb} needs to be computed.

\subsubsection{\Ac{rlrt}}

Differently from the \ac{glrt}, this test exploits the channel correlations among the \ac{ap} antennas. For deriving the test statistic, we follow the following procedure:
\begin{enumerate}
	\item We denote with $\mathbf{r}_m$, $m=1,2,...,N_{\textnormal{RE}}$ the column vector collecting the entries of $\mathbf{r}$ received by all antennas on RE $m$.
	\item We define $\mathbf{R} = [\mathbf{r}_1, \dots, \mathbf{r}_{N_{\textnormal{RE}}}]$, which is a $(N_{\textnormal{ant}} \times N_{\textnormal{RE}})$-dimensional matrix.
	\item We compute the sample covariance matrix as $\mathbf{C} = \frac{1}{N_{\textnormal{RE}}} \mathbf{R} \mathbf{R}^{\textnormal{H}}$.
	\item We define the test statistic as \cite{nadler_2011}
	\begin{equation}
	\label{eq:rlrt}
		T_{\textnormal{RLRT}} = \frac{\lambda}{\sigma_w^2} \, ,
	\end{equation}
	where $\lambda$ is the largest eigenvalue of $\mathbf{C}$.
\end{enumerate}
The threshold for this detector is derived as
\begin{equation}
	\label{eq:rlrt_approx}
	\delta_{\textnormal{RLRT}} \approx \mu + \xi \cdot F^{-1}_{\textnormal{TW2}} (1 - P_{\textnormal{FA}}) \, ,
\end{equation}
where $\textnormal{TW2}$ is the Tracy-Widom distribution of 2\textsuperscript{nd} order, while $\mu$ and $\xi$ depend on $N_{\textnormal{ant}}$ and $N_{\textnormal{RE}}$. In particular, authors in \cite{nadler_2011} show that the approximation holds for $N_{\textnormal{ant}}, N_{\textnormal{RE}} \rightarrow \infty$.

\begin{figure}
	\centering
	\includegraphics[width=\columnwidth]{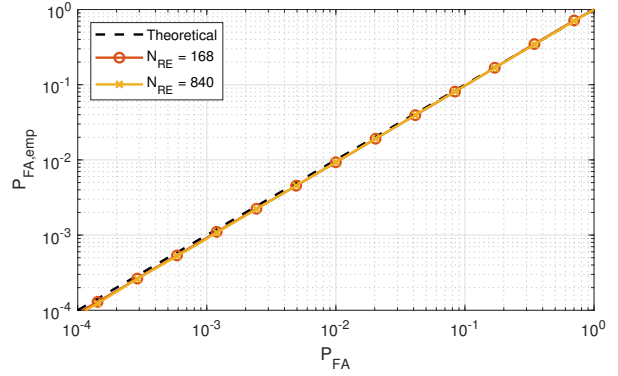}
	\caption{Empirical \ac{fa} probability versus target \ac{fa} probability for $N_{\textnormal{ant}} = 16$.}
	\label{fig:test_pfa_det_rlrt_ant16}
\end{figure}

When compared to the \ac{glrt}, with this detector we exploit the spatial correlation among antennas. The computational complexity increases, but is still very low as we just need to compute an eigenvalue. A second potential disadvantage is that the approximation in (\ref{eq:rlrt_approx}) creates a mismatch between empirical and target \ac{fa} probabilities. Therefore, in order to evaluate the impact of this mismatch, in Fig. \ref{fig:test_pfa_det_rlrt_ant16} we show the empirical \ac{fa} probability derived in an authentic scenario, i.e., a scenario without jamming, versus the target \ac{fa} probability, for a factory with $N_{\textnormal{ant}} = 16$ antennas. Three different curves are displayed: a theoretical one, for which the two probabilities coincide, and two empirical curves corresponding to $N_{\textnormal{RE}} = 168,840$ (i.e., $M_P = 1,5$). As we can see, both the empirical curves are close to the theoretical one, meaning that the approximation (\ref{eq:rlrt_approx}) holds very well even with realistic low values of $N_{\textnormal{ant}}$ and $N_{\textnormal{RE}}$.

\subsection{Jamming mitigation strategies}

Alongside the above detection strategy, we consider two jamming mitigation schemes designed for narrow-band attacks: one based on user scheduling and the other one exploiting the pseudo-random blanking concept.

In Section \ref{sec:sim_results} we will assume \emph{sequential scheduling} as baseline, such that adjacent \acp{prb} are allocated to each active UE. As a first mitigation strategy, we consider \emph{random scheduling}, where \acp{prb} are allocated to each \ac{ue} in a pseudo-random way, with the constraint that still, as introduced in Section \ref{subsec:numerology}, a \ac{prb} is allocated to just one active \ac{ue}, to guarantee orthogonality among \acp{ue}. The purpose of this approach is to counteract smart jammers that can learn allocation and, for instance, focus their attack on a specific subband that is used by just one or few \acp{ue}. With this method then the jammer cannot know in advance which \ac{ue} will be scheduled on each \ac{prb}.

As a second mitigation strategy, we consider \emph{frequency hopping}, where in each slot just a small number of \acp{prb} is used for transmission, and that is implemented in our framework by greatly increasing the number of blanked \acp{prb} $M_P$. The main objective is to lower the probability of intersection between jammed and data \acp{prb}, so advantages of frequency hopping are expected with narrow- rather than wide-band jammers. When using a large number of blanked \acp{prb}, the same packet needs to be transmitted on a lower number of data \acp{prb} but with higher power per \ac{prb}, i.e., a higher packet spectral efficiency is needed in (\ref{eq:bler}), but higher \ac{sinr} is also experienced on those data \acp{prb}: that, in fact, can be beneficial in certain interference conditions. Moreover, a large number of blanked \acp{prb} has the benefit of performing jamming detection on more resources, thus decreasing the \ac{md} probability.

\FloatBarrier

\section{Numerical Results}\label{sec:sim_results}

In this section we show the numerical results obtained by performing Monte Carlo simulations of the above described system. In particular, we focus on the system \acp{kpi} degradation caused by the jammer and on the \ac{md} probability of the attacker. If not otherwise specified, the following parameters are used for the simulations: $N_{\textnormal{ant}} = 64$ antennas, high power jammer with $P_\textnormal{J} = 60 \, \textnormal{dBm}$, $M_P = 5$ blanked \acp{prb}, and random scheduling of \acp{ue}.

\begin{figure}[t!]
	\centering
	\includegraphics[width=\columnwidth]{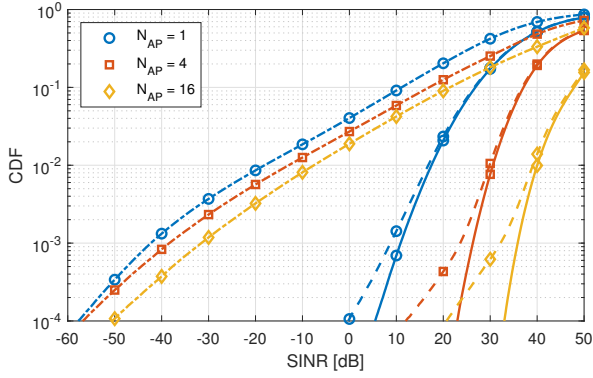}
	\caption{\ac{cdf} of \ac{sinr} for $B = 20 \, \textnormal{MHz}$ and $L_P = 25$. Continuous lines are without jamming, dashed lines for $P_\textnormal{J} = 20 \, \textnormal{dBm}$, and dash-dotted lines for $P_\textnormal{J} = 60 \, \textnormal{dBm}$.}
	\label{fig:CDF_SINR_PRB25_PRBJ25}
\end{figure}
\begin{figure}[t!]
	\centering
	\includegraphics[width=\columnwidth]{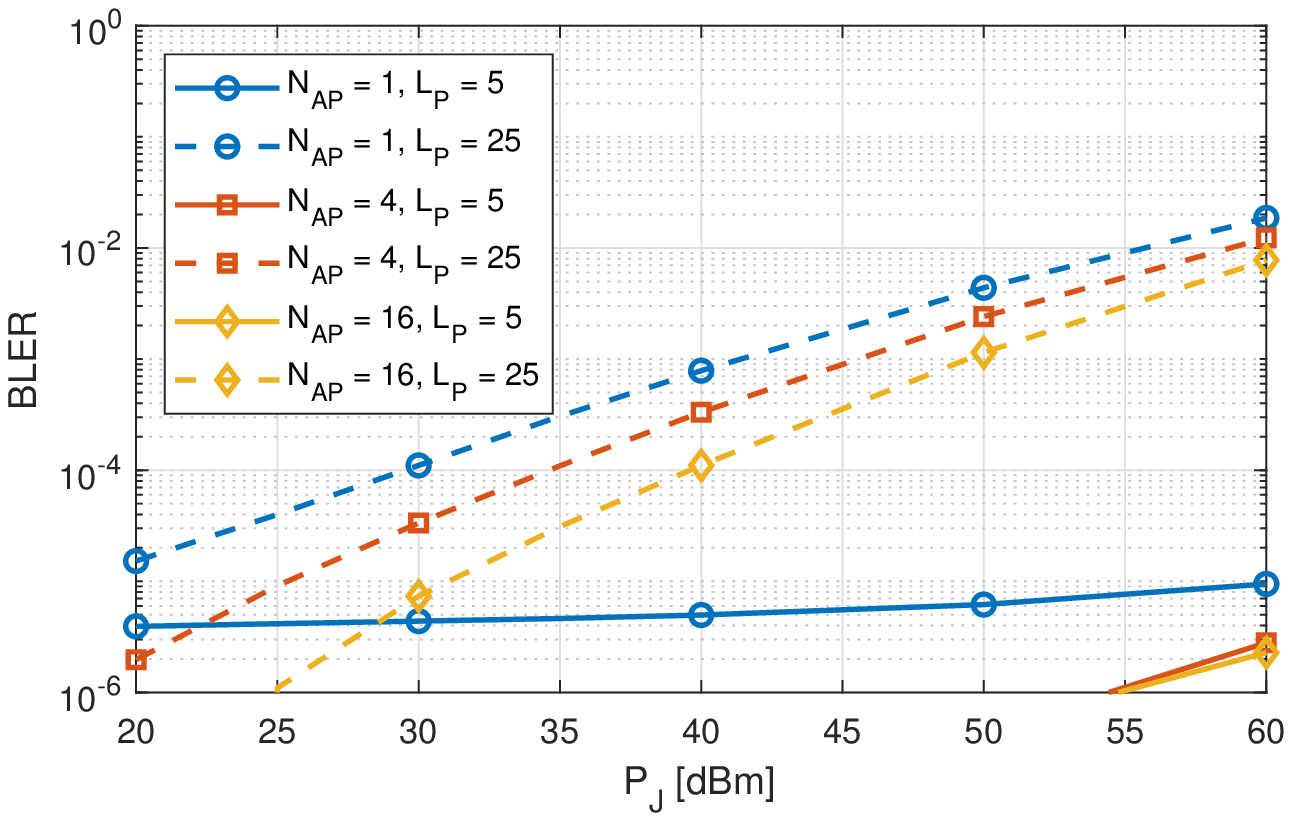}
	\caption{\ac{bler} versus $P_\textnormal{J}$ for $B = 20 \, \textnormal{MHz}$.}
	\label{fig:BLER_vs_PJ_B20_jammer_type_comparison}
\end{figure}
Fig. \ref{fig:CDF_SINR_PRB25_PRBJ25} shows the \ac{cdf} of the \ac{sinr} for $N_{\textnormal{AP}} = 1,4,16$, $B = 20 \, \textnormal{MHz}$, $P_\textnormal{J} = 20,60 \, \textnormal{dBm}$ (low- and high-power jammer), and $L_P = 25$ (wide-band jammer). Moreover, the \ac{snr} curves are also shown, representing a jamming free scenario. First, we notice as expected that the \ac{sinr} is higher in the distributed deployments, i.e., with higher $N_{\textnormal{AP}}$, because some of the \ac{ap} antennas are closer to the \acp{ue}. On the other hand, with jamming the \ac{sinr} gap among the deployments is reduced when compared to the jamming free scenario: that happens because some of the \ac{ap} antennas are, with the distributed approaches, also closer to the jammer stationed outside the factory. Finally, we observe that, while on the median the \ac{sinr} is still quite high even with a high-power jammer, on lower quantiles the \ac{sinr} is strongly affected, for instance with about $50 \, \textnormal{dB}$ loss at the 1st percentile, i.e., considering a \ac{cdf} value of 0.01, with $N_{\textnormal{AP}}=16$.

To evaluate the performance degradation with \ac{urllc} type of traffic, Fig. \ref{fig:BLER_vs_PJ_B20_jammer_type_comparison} shows the \ac{bler} (\ref{eq:bler}) as a function of $P_\textnormal{J}$ for $N_{\textnormal{AP}} = 1,4,16$, $B = 20 \, \textnormal{MHz}$, and $L_P = 5,25$ (narrow- and wide-band jammer). Better \ac{bler} is achieved by the distributed deployments. Moreover, we observe that the wide-band attack is much more harmful than the narrow-band attack, and a huge \ac{bler} degradation is observed with a wide-band jammer: for instance, \ac{bler} increases with $N_{\textnormal{AP}}=4$ from about $10^{-6}$ to $10^{-2}$ when we increase the jamming power from $20 \, \textnormal{dBm}$ to $60 \, \textnormal{dBm}$.

\begin{figure}[t!]
	\centering
	\includegraphics[width=\columnwidth]{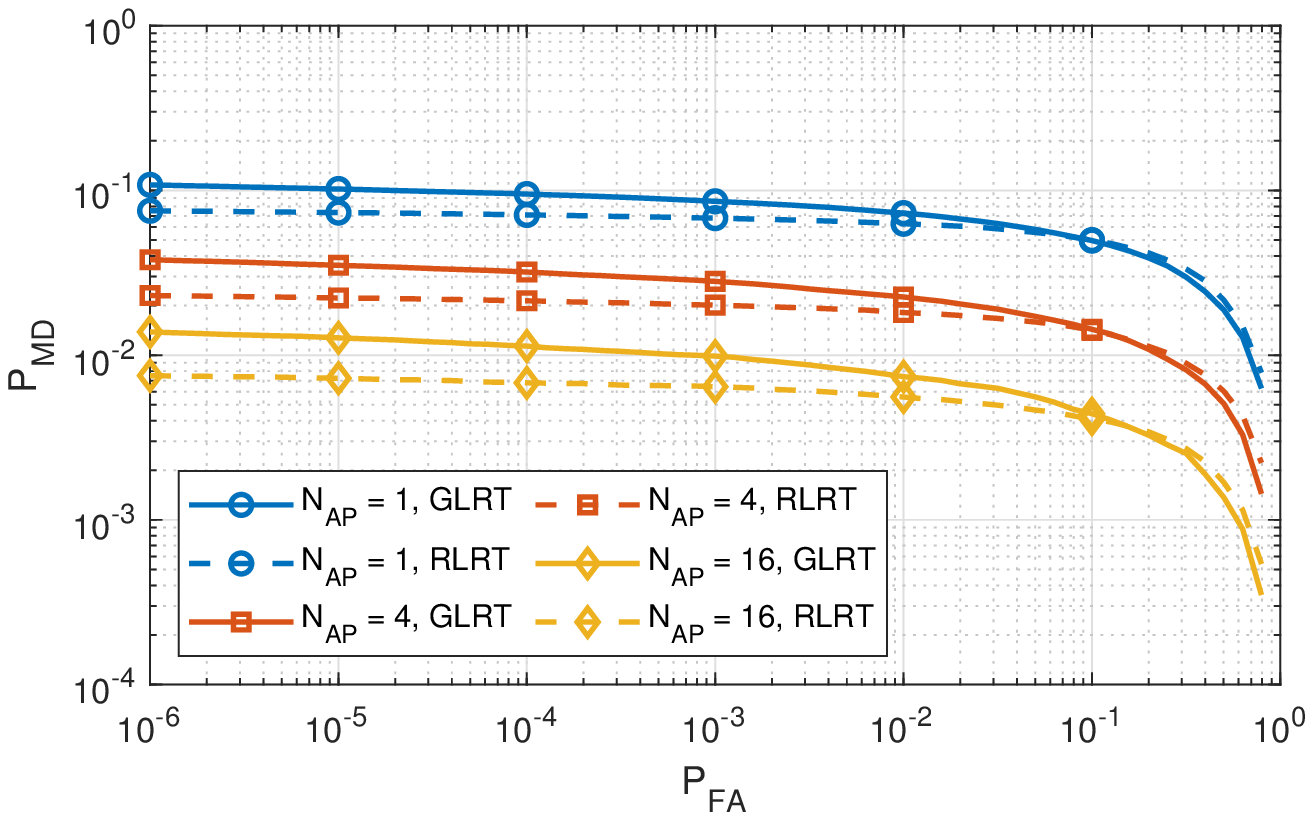}
	\caption{$P_\textnormal{MD}$ versus $P_\textnormal{FA}$ for $B = 20 \, \textnormal{MHz}$ and $L_P = 25$.}
	\label{fig:PMD_vs_PFA_B20_detectors_comparison}
\end{figure}

Regarding the performance evaluation of the defense strategy, Fig. \ref{fig:PMD_vs_PFA_B20_detectors_comparison} shows the \ac{md} probability as a function of the \ac{fa} probability, a.k.a. \ac{roc} curve, for $B = 20 \, \textnormal{MHz}$, $L_P = 25$, and comparing \ac{glrt} against \ac{rlrt} detectors. The first thing to notice is that the \ac{md} probability is lower, i.e., better, in the distributed approaches because \ac{ap} antennas are closer to the jammer. Then, \ac{md} probability is slightly lower with the \ac{rlrt} detector for relevant values of \ac{fa} probability, confirming that exploiting spatial correlation among antennas brings benefit to the detection.

\begin{figure}[t!]
	\centering
	\includegraphics[width=\columnwidth]{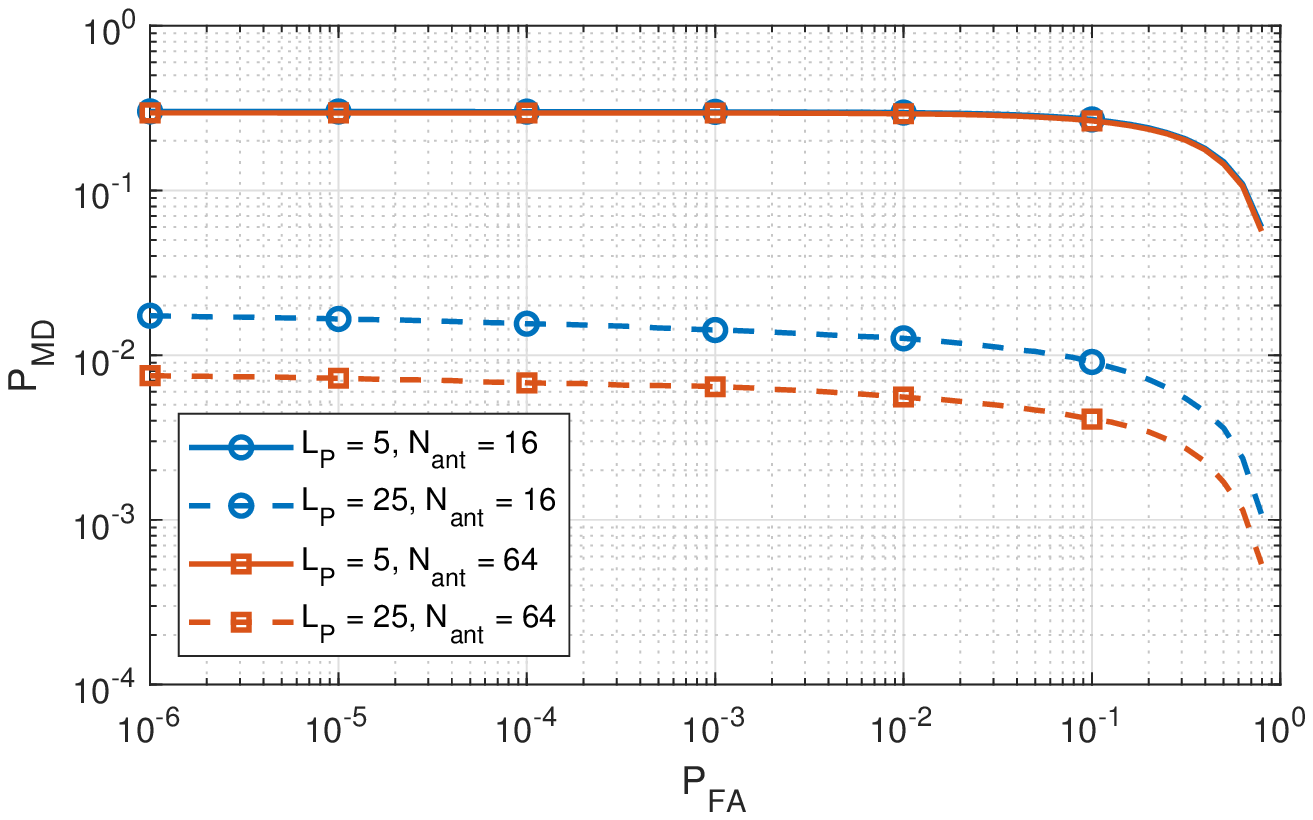}
	\caption{$P_\textnormal{MD}$ versus $P_\textnormal{FA}$ for $N_{\textnormal{AP}} = 16$, $B = 20 \, \textnormal{MHz}$, and \ac{rlrt} detector.}
	\label{fig:PMD_vs_PFA_B20_antenna_number_comparison}
\end{figure}

In Fig. \ref{fig:PMD_vs_PFA_B20_antenna_number_comparison} we show the \ac{roc} curve for $N_{\textnormal{AP}}=16$, $B = 20 \, \textnormal{MHz}$, $N_{\textnormal{ant}} = 16, 64$, $L_P = 5,25$, and \ac{rlrt} detector. Lower \ac{md} probability is achieved with more \ac{ap} antennas. On the other hand, in the narrow-band case \ac{md} probability is high and similar for different number of antennas, because limited by the probability of intersection between blanked and jammed \acp{prb}.

\begin{figure}
	\centering
	\includegraphics[width=\columnwidth]{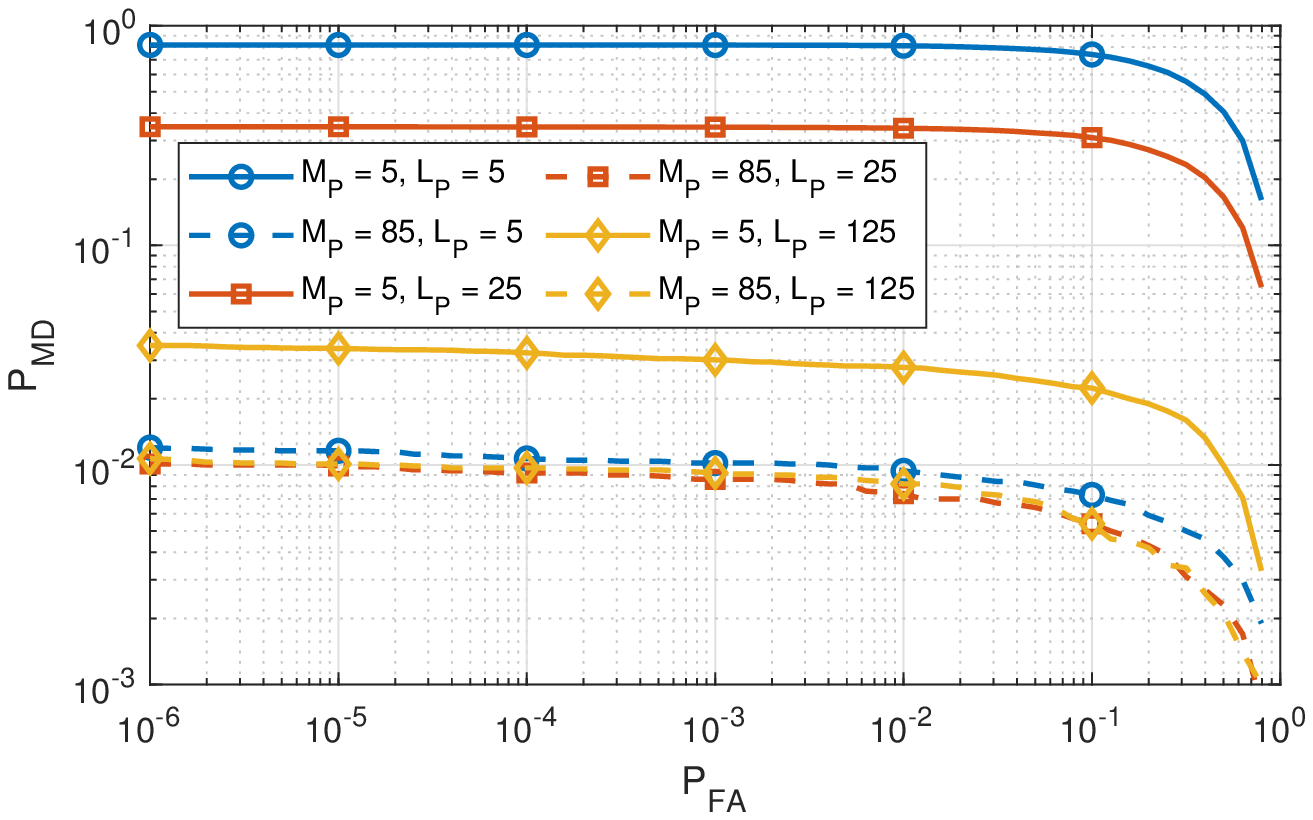}
	\caption{$P_\textnormal{MD}$ versus $P_\textnormal{FA}$ for $N_{\textnormal{AP}} = 16$, $B = 100 \, \textnormal{MHz}$, and \ac{rlrt} detector.}
	\label{fig:PMD_vs_PFA_B100_blanking_comparison}
\end{figure}

As last result regarding the detection performance, in Fig. \ref{fig:PMD_vs_PFA_B100_blanking_comparison} we report the \ac{roc} curve for $N_{\textnormal{AP}}=16$, $B = 100 \, \textnormal{MHz}$, $M_P = 5,85$, $L_P = 5,25,125$ (very narrow-band, narrow-band and wide-band jammer), and \ac{rlrt} detector. In this case, thanks to the larger number of available \acp{prb}, a massive blanking approach can be implemented and, indeed, \ac{md} probability is lower with more blanked \acp{prb}. Moreover, with massive blanking \ac{md} probability is similar across the different jamming strategies, thus allowing to better detect narrow-band jammers.

\begin{figure}
	\centering
	\includegraphics[width=\columnwidth]{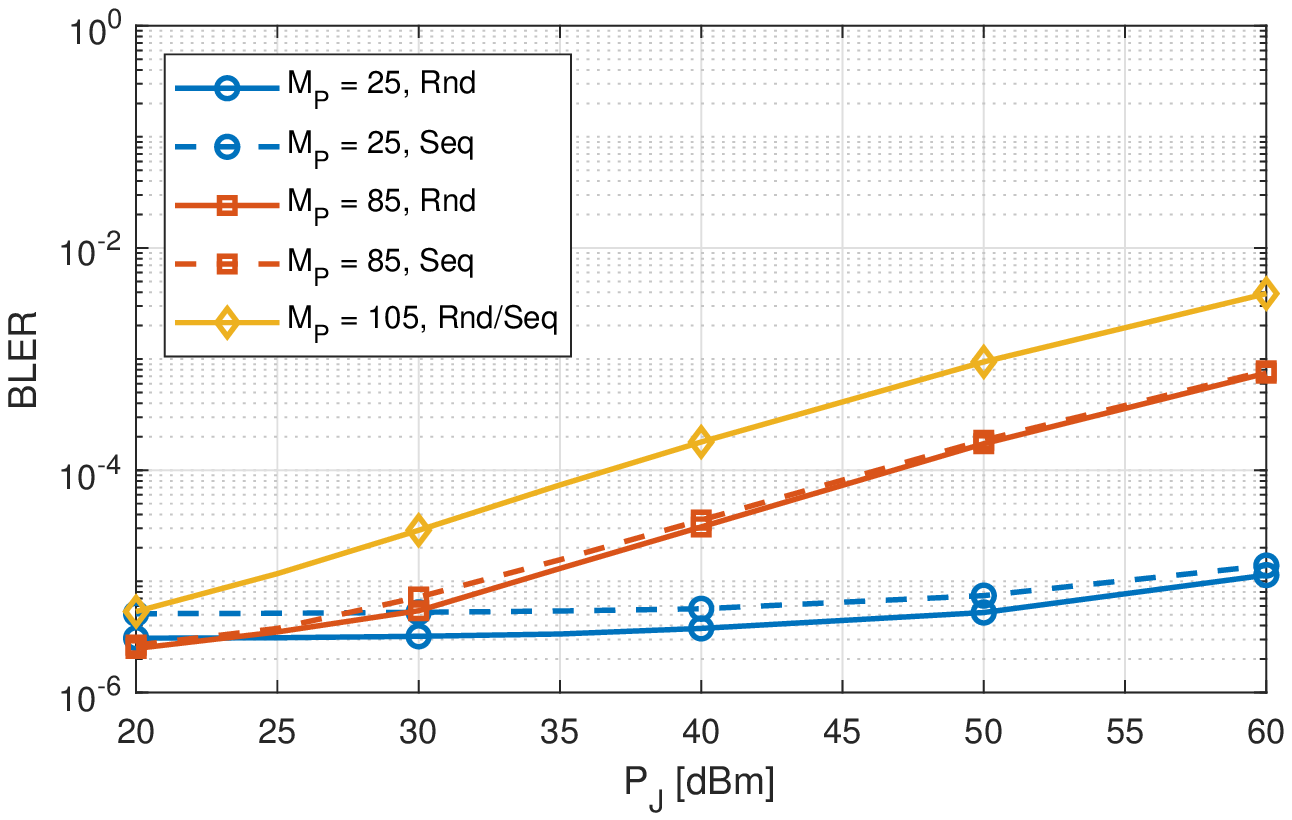}
	\caption{\ac{bler} versus $P_\textnormal{J}$ for $N_{\textnormal{AP}} = 1$, $B = 100 \, \textnormal{MHz}$, and $L_P = 25$.}
	\label{fig:BLER_vs_PJ_B100_PRBJ25_scheduling_comparison}
\end{figure}

Regarding the comparison among the different mitigation strategies, we consider Fig. \ref{fig:BLER_vs_PJ_B100_PRBJ25_scheduling_comparison}, which reports \ac{bler} as a function of $P_\textnormal{J}$ for for $N_{\textnormal{AP}} = 1$, $B = 100 \, \textnormal{MHz}$, random and sequential scheduling, $M_P = 25,85,105$, and $L_P = 25$. First, we notice that the scheduling-based mitigation works, although just a very small improvement is achieved by random scheduling when compared to the sequential one. Then, we observe a trade-off when applying frequency hopping: small $M_P$ (large bandwidth for data transmission) provides better performance in most ranges, but frequency hopping (large $M_P$) starts obtaining better performance when the jamming power is low, under whose conditions lower \ac{bler} can also be achieved by the system. In other words, these results tell that frequency hopping becomes helpful as a jamming mitigation scheme mainly when reliability requirements with \ac{urllc} are stricter, otherwise the increase in SINR is not sufficient to even compensate for the reduced bandwidth.
\section{Conclusions}\label{sec:conclusions}

In this paper we considered the problem of jamming attacks in \ac{5g}-and-beyond indoor factory deployments. We a) provided extensive simulations in a realistic scenario of a factory hall with \ac{3gpp} spatial channel model and a jammer stationed outside the plant, b) proposed and compared two detectors based on pseudo random blanking of subcarriers, and c) evaluated random scheduling and frequency hopping as jamming mitigation strategies. Numerical results show that a high-power jammer can strongly degrade \ac{bler} with \ac{urllc}. As promising countermeasures, a distributed deployment is more jamming resilient than a centralized one, and the \ac{rlrt} detector is capable to provide good jamming detection performance by exploiting channel correlations among the deployed antennas. Finally, frequency hopping is beneficial in mitigating jamming attacks only with narrow-band jammers and with more strict reliability requirements. Future works will include more advanced mitigation schemes exploiting \ac{mimo} and multi-connectivity.

\balance 

\bibliographystyle{IEEEtran}
\bibliography{IEEEabrv,bibliography}

\end{document}